\documentclass[openacc]{rsproca_new}%%%%where rsproca is the template name
\usepackage{orcidlink}
\usepackage{hyperref}
\usepackage{enumerate} % fancier enumeration options
\usepackage{multirow} % allow table cells spanning multiple rows
\usepackage{lmodern} % allow Latin Modern font family
\newcommand{\dma}[1]{\textcolor{red}{}}
\newcommand{\kut}[1]{\textcolor{blue}{}}
\newcommand{\Ne}{N_{\mathrm{e}}}
\newcommand{\Naug}{N_{\mathrm{aug}}}

\jname{rsif}
\Journal{J R Soc Interface\ }

\begin{document}

%%%% Article title to be placed here
\title{A new method for augmenting short time series, with application to pain events in sickle cell disease}

\author{%%%% Author details
Kumar Utkarsh \orcidlink{0000-0003-2552-1374}$^{1}$, Nirmish R. Shah \orcidlink{0000-0002-7506-0935}$^4$, Tanvi Banerjee\orcidlink{0000-0002-9794-3755}$^5$, Daniel M. Abrams \orcidlink{0000-0002-6015-8358}$^{1,2,3}$}

%%%%%%%%% Insert author address here
\address{$^{1}$Department of Engineering Sciences and Applied Mathematics, Northwestern University, Evanston, IL, USA\\
$^{2}$Northwestern Institute for Complex Systems, Northwestern University, Evanston, IL, USA\\
$^{3}$Department of Physics and Astronomy, Northwestern University, Evanston, IL, USA\\
$^{4}$Department of Medicine, Duke University, Durham, NC, USA\\
$^{5}$Department of Computer Science and Engineering, Wright State University, Dayton, OH, USA}

%%%% Subject entries to be placed here %%%%
\subject{computational biology, computational mathematics, healthcare modeling}

%%%% Keyword entries to be placed here %%%%
\keywords{sickle cell disease, stochastic process, Hawkes process, maximum likelihood estimation}

%%%% Insert corresponding author and its email address}
\corres{Kumar Utkarsh\\
\email{krutkarsh@u.northwestern.edu}}

%%%% Abstract text to be placed here %%%%%%%%%%%%
\begin{abstract}

    Researchers across different fields, including but not limited to ecology, biology, and healthcare, often face the challenge of sparse data. Such sparsity can lead to uncertainties, estimation difficulties, and potential biases in modeling. Here we introduce a novel data augmentation method that combines multiple sparse time series datasets when they share similar statistical properties, thereby improving parameter estimation and model selection reliability. We demonstrate the effectiveness of this approach through validation studies comparing Hawkes and Poisson processes, followed by application to subjective pain dynamics in patients with sickle cell disease (SCD), a condition affecting millions worldwide, particularly those of African, Mediterranean, Middle Eastern, and Indian descent.
     %present a point process model to describe 
     %Pain is a significant and one of the most debilitating symptoms of SCD, often occurring unexpectedly and varying in intensity. By better understanding pain dynamics, we aim to enhance patient care and tailor interventions more effectively. Facing similar data sparsity challenges as mentioned, we also introduce a method to combine multiple sparse datasets to circumvent some challenges posed by data sparsity. \dma{modify to emphasize method over application to SCD}
\end{abstract}
%%%%%%%%%%%%%%%%%%%%%%%%%%%

%%%%%%%%%% Insert the texts which can accomdate on firstpage in the tag "fmtext" %%%%%
\begin{fmtext}

%%%%%%%%%%%%%%% SECTION %%%%%%%%%%%%%%%%%%
\section{Introduction}
    
%Across fields of science and engineering, there is an essential need to account for complexity, randomness, and uncertainty. These elements are not merely theoretical, but they influence decision-making processes and outcomes in profound ways. When it comes to healthcare research, the significance of patient-reported data in providing insights into patient experiences and care quality has gained prominence \cite{valderas2008impact}. This data, however, is often characterized by inherent variability, including fluctuations in reporting frequency and content, which complicates analysis. Moreover, the temporal dynamics associated with health events, such as symptom fluctuations or treatment responses, pose a multitude of analytical challenges. Stochastic models have shown to be powerful tools for analyzing such data, allowing us to capture patterns and uncertainties of health events over time. 

Patient-reported data (including data from wearable devices) has gained increasing prominence in healthcare settings recently \cite{Pyper2023_DigitalHealthPatientGenerated, Nowell2023_DigitalTrackingRheumatoidRA, Kang2022_WearingTheFuture, Wettstein2024_RPMWearablePROM, Gagnon2024_WearableChronicSelfMgmt, Zheng2023_WearablesPostOpJointReplacement, hassan2025utility, vuong2023sickle, stojancic2023sickle}. This data, however, is often characterized by inherent variability, including fluctuations in reporting frequency and content, which complicates analysis. Moreover, the temporal dynamics associated with health events, including symptom fluctuations or treatment responses, pose a multitude of analytical challenges. Stochastic models provide tools for analyzing such data, when sufficiently abundant, allowing researchers and clinicians to learn useful information from observed patterns over time \cite{teng2020stocast, Kaplan2022_ContinuousTimeEHR, Liu2021_BayesianSTSensors}. However, attempting to draw conclusions from too narrow a set of observations can lead to overfitting, unreliable estimates, and difficulties in generalizing. \cite{Babyak2004_Overfitting, Ioannidis2005_WhyFalse, Riley2020_SampleSizePrediction}

\setlength{\parfillskip}{0pt} % justify last line of par
To address these issues, researchers have developed methodologies such as Bayesian approaches, which incorporate prior information and uncertainty \cite{gelman2013bayesian}, and hierarchical models that pool data across related groups to enhance estimates for conditions with sparse data \cite{raudenbush2002hierarchical}. Additionally, augmentation techniques like bootstrapping allow for the generation of synthetic data points, improving the robustness of statistical analyses \cite{efron1994bootstrap}. However, little has been 
\end{fmtext}
%%%%%%%%%%%%%%% End of first page %%%%%%%%%%%%%%%%%%%%%

\maketitle

\newpage

\noindent done on sparse sampling in the context of dynamical systems, with their inherent strong temporal correlation.  In this paper, we develop a new method that has broad applicability to sparsly sampled data from dynamical processes, and we focus in particular on testing the method in the real-world context of data capturing pain events in patients with sickle cell disease.

%\dma{Move this reference elsewhere or delete: These challenges and solutions are particularly pertinent in healthcare, where patient-reported data is often sporadic and subject to inherent variability \cite{valderas2008impact}.}

%This paper will focus on a stochastic modeling approach, specifically employing point process methods, to analyze the dynamics of pain episodes in SCD patients. Statistical models, particularly stochastic approaches like point processes, are especially useful for capturing the randomness and temporal patterns of pain events. In contrast to differential equations models, which may require precise parameterization of underlying biological processes, or machine learning models that often demand large datasets and may overlook temporal dependencies, stochastic models provide a flexible framework that can adapt to the inherent variability of pain episodes in SCD patients.

%%%%%%%%%%%%%%% SECTION %%%%%%%%%%%%%%%%%%
\section{Model and Methods}

\subsection{Model and Assumptions}

%%%%%%%%%%%%%%%%%%%%%%%%%%%%%%%%%
\begin{figure}[t]
    \centering
    \includegraphics[width=0.55\columnwidth]{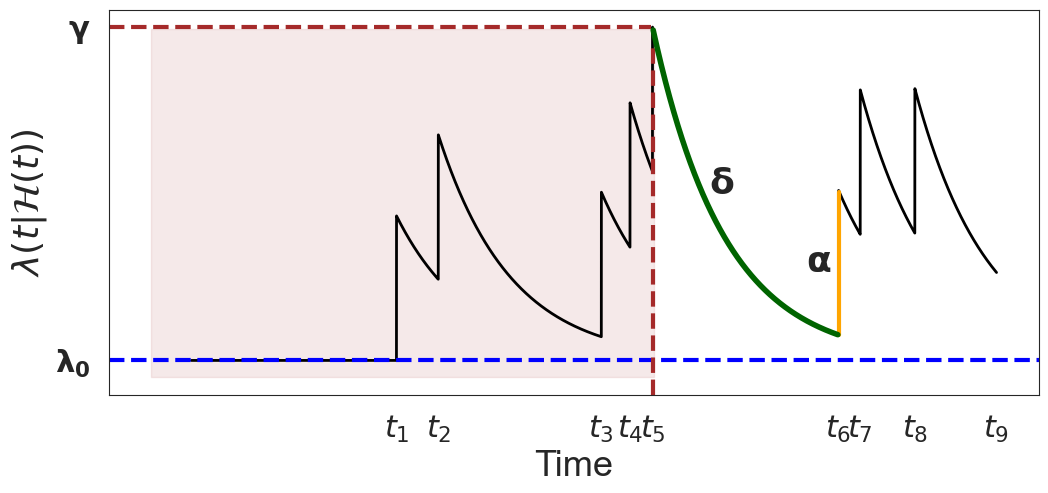}
    \caption{\textbf{Visual guide to shifted Hawkes process parameters and intensity dynamics.} Characterization of the parameters introduced in Eq.~\eqref{shiftedhawkes} (see also Table \ref{table:hawkesparams}). The peaks represent event arrivals in real-time. The shaded area represents the history not captured in the observed data. In this example, the observable $t=0$ is $t_5$. The blue horizontal dashed line shows the value of the baseline intensity $\lambda_0$ and the red horizontal dashed line shows the initial intensity value $\gamma$. The amplitude of impact $\alpha$ is depicted by the length of the vertical yellow line segment, whereas the decay rate $\delta$ characterizes the exponential decay curves, like the one shown in green.}
    \label{fig:shiftedhawkesparams}
\end{figure}
%%%%%%%%%%%%%%%%%%%%%%%%%%%%%%%%%

%%%%%%%%%%%%%%%%%%%%%%%%%%%%%%%%%
\begin{table}[t]
    \centering
    \begin{tabular}{ |c|p{12cm}|c| } 
        \hline
        \textbf{Parameter} & \textbf{Description} & \textbf{Units} \\
        \hline
        \hline
        %\hline
        $\lambda_0$ & Baseline intensity value for underlying homogeneous Poisson process. & $[T^{-1}]$ \\ 
        \hline
        $\alpha$ & Amplitude of impact of an individual event arrival on intensity. & $[T^{-1}]$\\ 
        \hline
        $\delta$ & Rate of decay to baseline intensity ($\delta^{-1}$ sets memory length). & $[T^{-1}]$\\ 
        \hline
        $\gamma$ & The intensity measurement recorded at the initiation of the data collection period. & $[T^{-1}]$ \\ 
        \hline
    \end{tabular}
    \caption{Description of model parameters introduced in Eq.~\eqref{shiftedhawkes} (see also Fig.~\ref{fig:shiftedhawkesparams}).}
    \label{table:hawkesparams}
\end{table}
%%%%%%%%%%%%%%%%%%%%%%%%%%%%%%%%%

Our approach is motivated by collections of datasets that have irregular sampling and limited length, but where at least a subset of the collection may be well explained by a single model.  This is true for our example collection of pain events in patients with SCD. 

In particular, the frequency of early hospital readmission in SCD patients (almost 90\% readmitted within 30 days \cite{adesina2023all,shah2019sickle}) suggests that  pain events may naturally cluster temporally. Therefore, we treat the occurrence of pain crisis events as a self-exciting process, meaning that patients are at elevated risk for a subsequent event immediately after one occurs. This type of process was first described mathematically by Alan G. Hawkes in the context of modeling seismic activity \cite{hawkes1971spectra, hawkes1974cluster, ogata1988statistical}, where earthquakes trigger aftershocks. The analogy extends naturally to SCD, where initial obstructions in blood flow result in inflammatory responses that may precipitate subsequent pain crises. Similar dynamics have been reported in criminal activities \cite{mohler2011self} and financial markets \cite{bowsher2007modelling}.  We thus treat the occurrence of pain crisis events as a \textit{Hawkes process}.

A Hawkes process is a counting process $\{N(t)|t\geq0\}$ with an associated history $\mathcal{H}(t):=\{t_i|t_i<t\}$, where $t_i$ is the time of $i^{th}$ event, and a conditional intensity function $\lambda(t|\mathcal{H}(t))$ of the form \begin{equation}
    \label{hawkescond1}
        \lambda(t|\mathcal{H}(t)) = \lambda_0(t) + \sum_{i:t>t_i}\Phi(t-t_i),
    \end{equation}
where $\lambda_0>0$ is the \textit{baseline intensity} and $\Phi\geq0$ is a monotonically decreasing function referred as the \textit{memory kernel}. 
%While our focus is on self-exciting processes where $\Phi \geq 0$, we note that self-inhibiting processes (with $\Phi < 0$) are possible in other contexts. 
In such a process, the occurrence of an event increases the likelihood of a subsequent event for some time after the initial arrival (with this ``memory timescale'' set by the decay rate of $\Phi$). Henceforth, the terms ``conditional intensity function'' and ``intensity function'' will be used interchangeably, and $\lambda(t)\equiv\lambda(t|\mathcal{H}(t))$.

We further simplify the Hawkes process model by assuming a constant baseline intensity $\lambda_0(t) = \lambda_0$ and an exponential memory kernel $\Phi(t-t_i) = \exp[-\delta (t-t_i)]$ with $\delta > 0$.  We choose the exponential kernel for three reasons: (a) it provides analytical tractability, allowing closed-form likelihood computation; (b) it gives clear parameter interpretability with $\delta^{-1}$ representing a characteristic memory timescale; and (c) it is physiologically plausible, as inflammatory cascades likely exhibit exponential decay kinetics.

One common assumption in the use of Hawkes models is that the process is observed from its onset. However, in practice, data may be collected during an arbitrary interval in the middle a longer process---this is the case for our SCD datasets, which begin months to years after the onset of SCD symptoms. Keeping that in mind, we introduce an additional compensatory term to capture past events not recorded in the dataset, $(\gamma-\lambda_0)e^{-\delta t}$.  This yields our proposed model intensity function
\begin{equation}
    \lambda(t) = \lambda_0 + \alpha \sum_{i:t>t_i} e^{-\delta (t-t_i)} + (\gamma-\lambda_0)e^{-\delta t},
    \label{shiftedhawkes}
\end{equation}
where $\lambda_0>0, \alpha\geq0, \delta>0,$ and $\gamma\geq0$ are constants and $t_i$ is the time of $i^{th}$ event. A description of each parameter can be found in Table \ref{table:hawkesparams} and parameters are visualized in Fig.~\ref{fig:shiftedhawkesparams}. 

We note that this model is designed for time series datasets that capture an event at $t=0$, which serves as the reference point for the initial condition $\gamma$.  The key parameters dictating model behavior are $\lambda_0$, $\alpha$ and $\delta$, whereas $\gamma$ should be seen as a correction factor to compensate for the effects of past history not captured in the dataset.

\subsection{Model Fitting and Selection}

%%%%%%%%%%%%%%%%%%%%%%%%%%%%%%%%%
\begin{figure}[t]
    \centering
    \includegraphics[width=0.6\columnwidth]{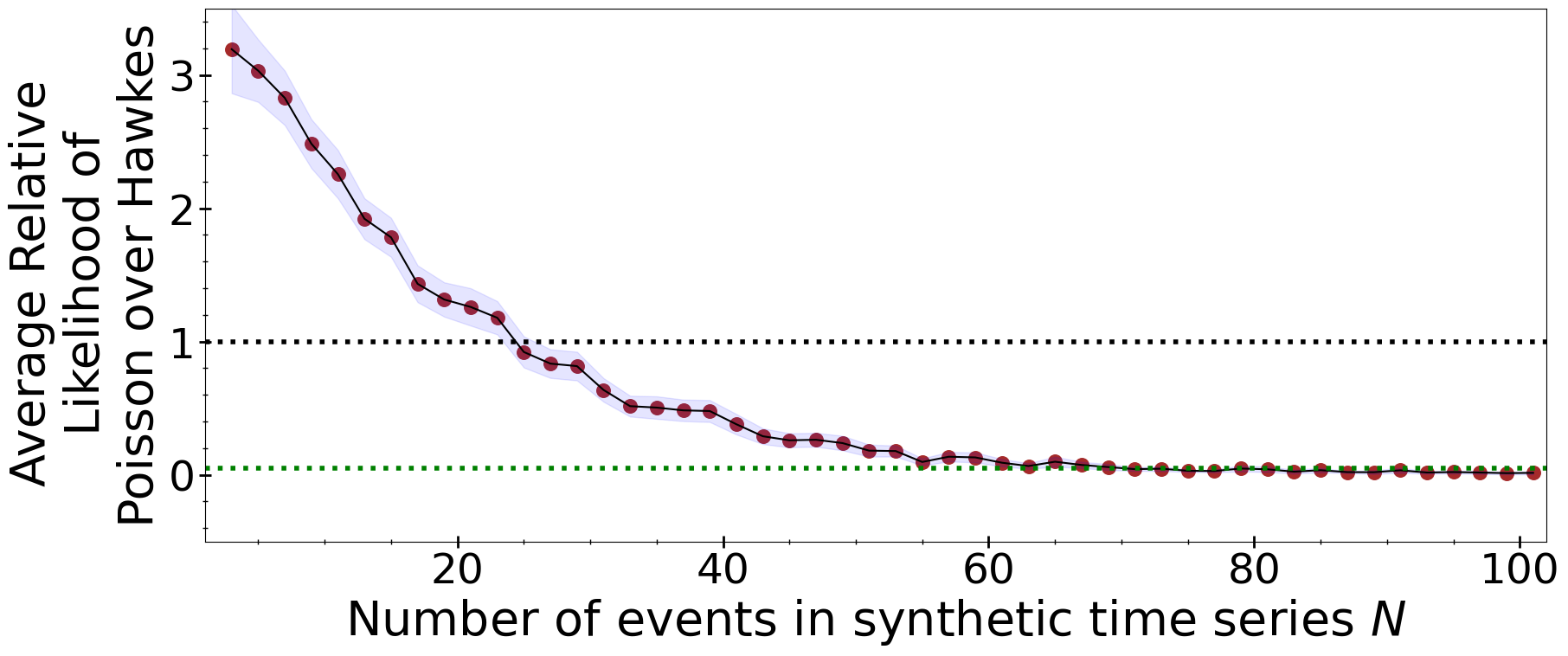}
    \caption{\textbf{Minimum dataset size required for reliable Hawkes vs. Poisson model discrimination.} Number of data points needed to distinguish Hawkes model from Poisson. The black dashed line is for basic preference ($\mathcal{L}=1$), whereas the green dashed line is for 95\% confidence ($\mathcal{L}=0.05$). For each $N$, we calculate the relative likelihood 50 times. The averages are calculated, and denoted by the red markers. The purple-shaded region denotes the 95\% confidence interval. The likelihood values and critical $N$ depend on the parametric choices. We use ($\lambda_0$, $\alpha$, $\delta$)=(1, 3, 6). %\dma{improve for publication: long run time version, make version for 1 column width (about 78mm)}
    }
    \label{fig:modelcomp}
\end{figure}
%%%%%%%%%%%%%%%%%%%%%%%%%%%%%%%%%

Before fitting to data, we wish to establish what would characterize a ``successful'' model.  In particular, we would like to choose a reasonable null model against which to test, and we hope to deduce exactly how much data is necessary to distinguish among models.

Substituting $\alpha=0$ and $\gamma=\lambda_0$ into Eq.~\eqref{shiftedhawkes} yields a homogeneous Poisson process (henceforth referred to simply as ``a Poisson process''), but this is just a special case of the Hawkes model where memory effects are absent.\footnote{The same limiting behavior arises when $\delta \to \infty$, meaning memories disappear instantaneously after each event.} Conversely, the Hawkes model can be seen as an extension over a baseline Poisson model with a self-exciting memory kernel. For this reason, the Poisson process is a natural candidate for a null model. In addition to the fact that it represents a simple and widely used framework for modeling point processes, we can assess whether the added complexity of the Hawkes process results in a statistically significant improvement.

We fit model parameters using maximum likelihood estimates, where likelihood for a dataset $\{t_i\}_{i=1}^N$ is given by
\begin{equation}
    \label{likelihood}
    L(\mathbf{\theta}|t_1, t_2,...,t_N) = \left(\prod_{i=1}^N \lambda(t_i)\right) e^{-\int_0^{t_N}\lambda(s) ds}.
\end{equation}
% For simplicity, we compare the Poisson process null model to the Hawkes model with complete history. 
For the Hawkes model, if we know the complete history (as in simulations), the process starts at baseline intensity $\gamma=\lambda_0$, and is defined by
\begin{equation}
    \label{hawkes}
    \lambda(t)= \lambda_0 + \alpha \sum_{i:t>t_i} e^{-\delta (t-t_i)}.
\end{equation}

We use the Akaike Information Criterion (AIC) for the comparison, as it provides a theoretically grounded balance between a model's goodness of fit and complexity \cite{akaike1974new} (though we note that there are potential problems with use of information criteria in dynamical systems \cite{utkarsh2025AIC}).  Our candidate model has four degrees of freedom ($\lambda_0, \alpha, \delta, \gamma$), while the null model has just one ($\lambda_\textrm{P}$). Since these models are nested, with the Poisson model being a simpler version of the Hawkes model, differences in complexity can make it challenging to recover the true model, especially with limited data. 

To illustrate, we simulate multiple time series of varying lengths using Eq.~\eqref{hawkes}. Fig.~\ref{fig:modelcomp} shows that AIC often reflects greater evidence for the Poisson model for shorter time series, indicating a threshold length (dependent on parameter choices) below which this approach fails to detect the true dependency structure. 
%
%Since the time series were generated using the Hawkes model, we know that the true process includes event dependence, which the Poisson model does not capture. The support for the Poisson model for smaller datasets, however, is due to insufficient evidence accumulation for the more complex Hawkes process. 
This aligns with the principle that AIC generally performs better with larger sample sizes, where the trade-off between fit and complexity becomes clearer \cite{hurvich1989regression}.\footnote{A version of Fig.~\ref{fig:modelcomp} using the corrected AIC (AICc) \cite{Hurvich1989} is provided in Appendix~\ref{correctedAICmodelcomp}. Corrections for finite data do not resolve the sparse data problem, they have the opposite impact of making it a greater challenge.}

%\dma{cut this paragraph but mention AICc somewhere}\dma{we should cite AICc and probably use that or show its implications on a graph like Fig 3 also---maybe in appendix or SI}\kut{Made changes but not deleted yet}
%Given the challenge of using information criteria with small, sparse datasets, it is crucial to explore methods that can enhance model selection reliability in such contexts. Traditional criteria, even with corrections, struggle to reliably capture model complexity in limited data settings. A version of Fig.~\ref{fig:modelcomp} using the corrected AIC (AICc) \cite{Hurvich1989} is provided in Appendix~\ref{correctedAICmodelcomp}, illustrating the impact of small-sample corrections. Therefore, developing approaches that augment sparse datasets presents a promising direction.

\subsection{Data Augmentation}

%\kut{Candidate acronyms: PAM (patient augmentation method), MERGE (Multi-source Ensemble for Robust Generalization and Estimation), MOSAIC (Model Optimization by Statistical Aggregation of Independent Collections), GLUE (Grouped Learning Using Ensembling)}

Our idea for augmenting sparse datasets is as follows: (1) we test datasets, pairwise, for statistical similarity; (2) we replace each individual dataset with an ensemble of those shown to be similar; and (3) we fit model parameters to each full ensemble, treating it as composed of disconnected excerpts drawn from a single process.

Concretely, we use the two-sample Kolmogorov-Smirnov (KS) test to identify datasets with similar distributions. The KS test is a non-parametric method that compares the cumulative distributions of two independent samples, quantifying the distance between them. This makes it an effective tool for identifying similarities within a collection, as it does not rely on assumptions about the underlying distributions and is sensitive to differences in both location and shape. 

\begin{figure}[t]
    \centering
    \includegraphics[width=0.3\linewidth]{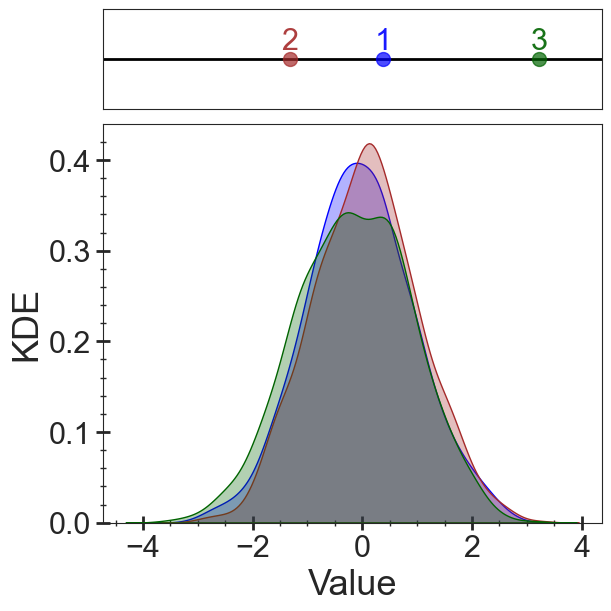}
    \hspace{5mm}
    \includegraphics[width=0.389\linewidth]{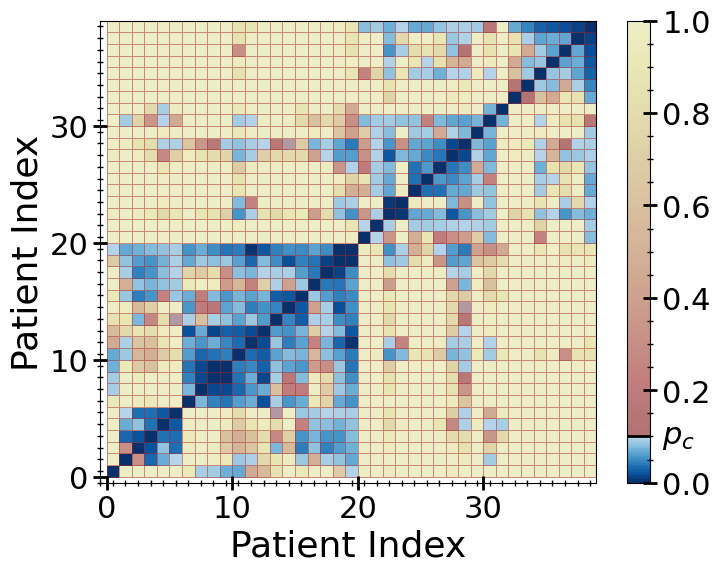} 
    \caption{\textbf{KS test demonstrates non-transitive similarity and identifies matchable patient pairs.} (\textit{Left}) Two-sample KS test for three normal distributions (1, 2, and 3) with slightly different means and variances (colored blue, red, and green, respectively). The three $p$-values for (1 vs.~2), (1 vs.~3), and (2 vs.~3) are 0.03, 0.02, and $10^{-5}$, respectively. This example demonstrates the non-transitive nature of the test: the mutual distances between the dots in the upper panel are in proportion to the mutual KS statistics; 1 is similar to both 2 and 3, whereas 2 and 3 are far apart and are thus dissimilar. (\textit{Right}) Similar patients in our collection of datasets using the two-sample KS test (0.1 significance level). Blue shades for matched pairs of patients and yellow-pink shades for unmatched.}
    \label{fig:kstest}
\end{figure}

By applying the KS test, we can systematically group datasets with comparable characteristics, enabling a collective analysis. The null hypothesis for this test is that two samples come from the same distribution. The test statistics and sample sizes are used to calculate the \textit{p}-values for the test, and the hypothesis is accepted if this \textit{p}-value exceeds our set threshold $p_c$. An example of its application is shown in Fig.~\ref{fig:kstest}.

We note that, since the KS test compares samples drawn from distributions, it cannot be applied directly to entire time series. To address this in the context of our patient data, we instead compare the distributions of interarrival times between events, which capture the underlying temporal structure. 

%%%%%%%%%%%%%%%%%%%%%%%%%%%%%%%%%%%%
%%%%%%%%%%%%%%%%%%%%%%%%%%%%%%%%%%%%

Once similar datasets are identified, we define a ``collective likelihood'' that integrates information across these matched groups, enhancing the reliability of model selection in contexts where individual datasets are too sparse for robust analysis. The following provides a step-by-step description of this strategy, applied to a collection of small or sparse time series where model selection might otherwise be unreliable:
\vspace{-\baselineskip}
\begin{enumerate}[\textit{Step} 1:]
    \item Consider a collection of $m$ time series datasets  $\mathcal{C} = \{t_i^j:i=1\ldots n_j, j=1\ldots m\}$. Compute the set of interarrival times $\Delta \mathcal{C} = \{\Delta t_i^j:i=1\ldots n_j-1, j=1\ldots m\}$. 
    %We denote the array as vectors $\mathbf{t^j}=\{t_i^j:1\leq i\leq n_j\}$.
    \item Calculate the $p$-values using a two-sample KS test for each pair of datasets of interarrival times and define the matrix $\mathbf{P} = \{p(i,j):i,j=1\ldots m\}$.  Those with $p$-values below a given threshold $p_c$ are taken to be similar.
    \item 
    %For any index $j$, let the indices corresponding to similar distributions be $\{j_1,j_2,...,j_k\}$. 
    We define the collective likelihood ($\tilde{L}$) of a model for dataset $j$ as 
    \begin{equation}
        \label{collectivelikelihood}
        \tilde{L}_j = L_j \prod_{i: \mathbf{P}_{ij} < p_c, i \neq j} L_{i},
    \end{equation}
    where $L_i$ is the individual likelihood of a given model for dataset $i$ and $p_c$ is a similarity threshold. This way, we consider the similar datasets to be different realizations of the same process. 
    \item We calculate the best-fit parameters and AIC values for each model using the collective likelihood. We use these for model selection.
\end{enumerate}
\vspace{-\baselineskip}
Note that this aggregation approach is non-transitive. If datasets A and B are statistically similar, and B and C are as well, it need not be true that A and C are.  Thus, the model parameters ultimately fitted to A, B, and C may all end up different. 

% With the data augmentation strategy established, we now turn to an analysis of its effectiveness across two distinct test cases. These cases are designed to evaluate how well the augmented datasets support reliable model recovery under conditions where individual datasets lack sufficient data. By examining the outcomes across these cases, we can assess the impact of grouping similar datasets through interarrival time distributions and using the collective likelihood in the patient dataset. 
%The following sections detail each test case, focusing on model recovery accuracy and the robustness of the augmented datasets in identifying the true data-generating process.

%%%%%%%%%%%%%% SECTION %%%%%%%%%%%%%%%%%%%
\section{Model Analysis and Method Verification}

% \dma{Move these two intro paragraphs somewhere else}

% The mean intensity of a stationary exponential Hawkes process \cite{laub2021elements} is
% \begin{equation}
%     \mathbb{E}[\lambda(t)]
%     = \frac{\lambda_0}{1 - \alpha/\delta}
%     \label{eq:stationaryhawkesrate}
% \end{equation}
% (see Appendix~\ref{app:hawkes_equil} for a brief derivation). 
% In addition to the individual contributions of the parameters, the ``branching ratio'' $\alpha/\delta$ plays a central role in Hawkes model fitting. In our experiments, the estimated parameters typically converge near---but not exactly to—their true values; in particular, the likelihood surface is notably flat in the $\alpha$--$\delta$ direction near the optimum (see Appendix~\ref{app:identifiability}). This subtlety particularly plays a role in the second subsection.

% Our findings from both the following test cases suggest that the data augmentation strategy not only preserves estimation accuracy but also offers a viable approach to improving model robustness when applied to small, sparse datasets. This approach could be especially valuable in domains where collecting continuous, high-frequency data is challenging, yet effective modeling of sporadic events is critical.

\subsection{Differentiating Hawkes from Poisson}

We consider two processes, a Poisson process with $\lambda_\textrm{P}=2$ and a Hawkes process (Eq.~\eqref{hawkes}) with comparable stationary behavior. Specifically we choose parameters such that the expected value of its equilibrium intensity $\mathbb{E}[\lambda(t)] = \lambda_0 (1 - \alpha/\delta)^{-1}$ is 2 as well (see Appendix \ref{app:hawkes_equil} for a brief derivation of this formula and Appendix \ref{app:identifiability} for a brief discussion of parameter identifiability issues).

The goal of this numerical experiment is to test whether the proposed data augmentation strategy improves model selection by enhancing the ability to distinguish between the two processes. Specifically, we evaluate the AIC differences between the Poisson and Hawkes models fitted to $\Naug=10$ original single-series datasets with $\Ne=30$ events each and a single augmented datasets for each model class.

%\dma{also interesting would be a test where the algorithm needs to decide which datasets belongs to which model.}
 
%%%%%%%%%%%%%%%%%%%%%%%%%%%%%%%%%%%%
\begin{figure}[t]
    \centering
    \includegraphics[width=45mm]{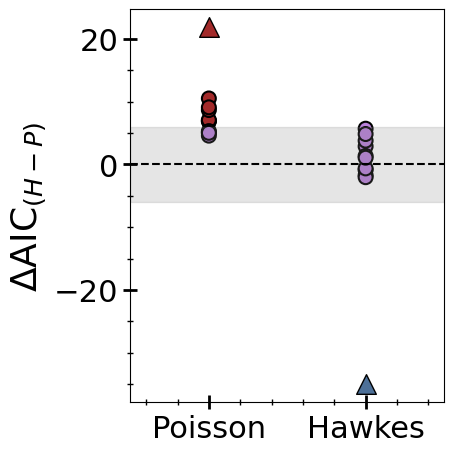}
    \caption{\textbf{Data augmentation shifts preference from inconclusive to confident model selection.} We generate 20 synthetic times series, 10 from a Poisson process and 10 from a Hawkes process, then examine the relative statistical support $\Delta\mathrm{AIC}_{(H-P)} = \mathrm{AIC}(H) - \mathrm{AIC}(P)$ for each model before and after augmentation. Each single-series realization is shown as a colored circle, and augmented datasets are shown as larger triangles with the same color scheme. Blue markers indicate a preference for the Hawkes model, red markers indicate a preference for the Poisson model, and purple markers correspond to inconclusive cases. Points above the dashed line correspond to a preference for the Poisson model, while points below correspond to a preference for the Hawkes model. Here $\Naug = 10$ and $\Ne = 30$ (events per time series). Regions outside the grey band correspond to model support with $>95\%$ confidence.  Parameters: ($\lambda_0=1$, $\alpha=2$, $\delta=3.5$) and $\lambda_P = 7/3$. }
    \label{fig:AIC_comparison}
\end{figure}
%%%%%%%%%%%%%%%%%%%%%%%%%%%%%%%%%%%%

The results are summarized in Fig.~\ref{fig:AIC_comparison}, where $\Delta$AIC values are shown for single-series (colored circles) and augmented datasets (larger triangles). Blue markers indicate a preference for the Hawkes model, red markers indicate a preference for the Poisson model, and purple markers correspond to inconclusive cases. The gray shading in the figure represents the inconclusive region, corresponding to results within the 95\% confidence interval where neither model is clearly preferred. The boundary for this inconclusive region can be derived from the relationship between $\Delta$AIC and relative likelihoods $\mathcal{L}_{\rm relative} = \exp \left(-\Delta \textrm{AIC}/2 \right)$. At the 95\% confidence level this yields $\Delta \textrm{AIC}^{\textrm{(crit)}} = -2\ln(0.05)\approx 6$ (so any $|\Delta \rm AIC| \gtrsim 6$ lies outside the inconclusive region and indicates strong evidence favoring one model over the other).

Augmenting datasets consistently moves results outside the inconclusive region for both processes, demonstrating that the strategy does indeed enable more robust model selection. 

% \dma{cut this?}For the Poisson-simulated datasets, the Hawkes model overfits due to its additional parameters, resulting in a higher AIC. Conversely, for the Hawkes-simulated datasets, the Poisson model fails to capture the temporal dependencies, leading to a significantly higher AIC.

%These findings validate the utility of the data augmentation strategy for improving model selection in stochastic processes with differing temporal dynamics. By amplifying the defining features of each process, the augmentation enhances the ability to distinguish between competing models.

\subsection{Parameter Estimation for Augmented Datasets}

%%%%%%%%%%%%%%%%%%%%%%%%%%%%%%%%%%%%
\begin{figure}[t]
    \centering
    \includegraphics[width=0.2\columnwidth]{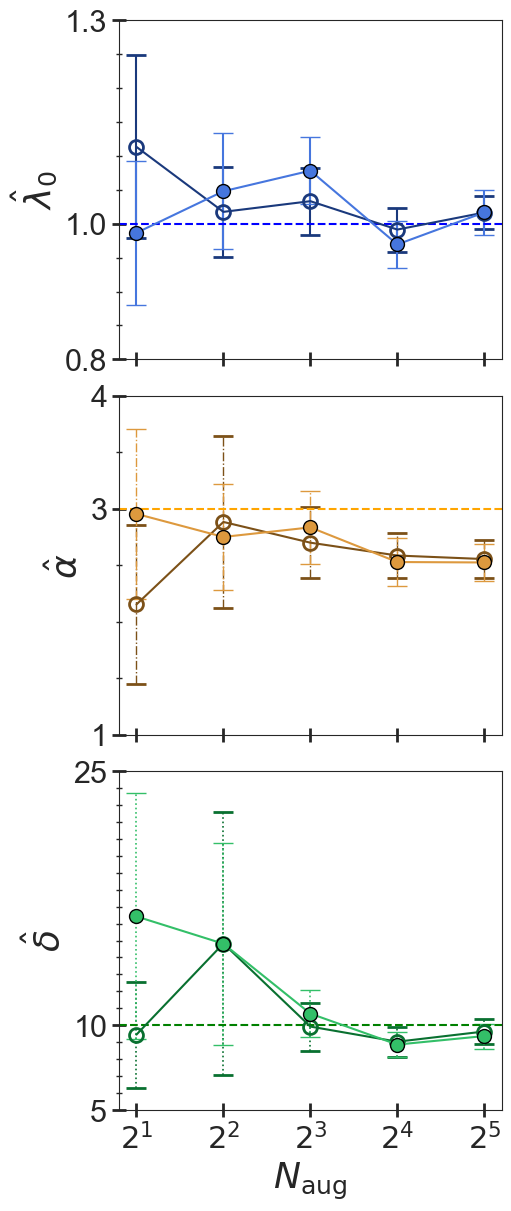} 
    \includegraphics[width=0.2\columnwidth]{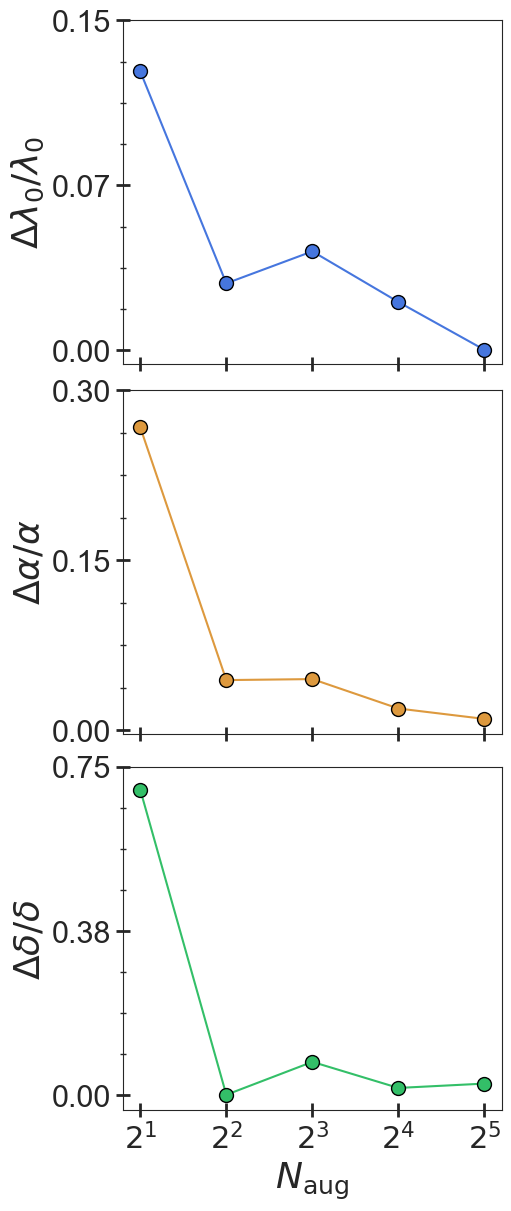} 
    \includegraphics[width=0.2\columnwidth]{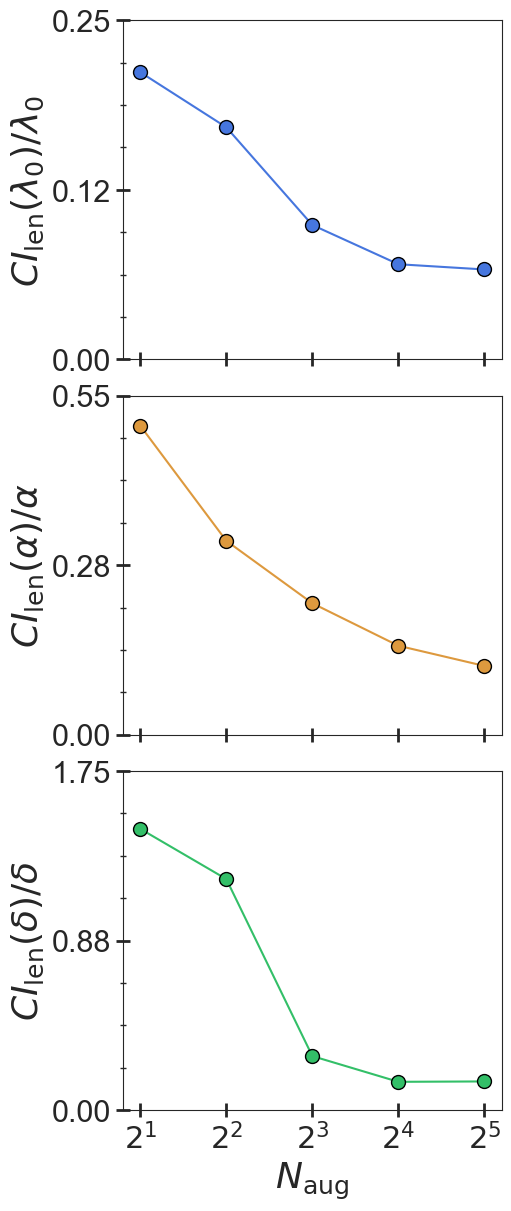} 
    \caption{\textbf{Augmented sparse series recover parameters comparably to equivalent-length continuous data.} Comparison of parameter estimation performance between our data augmentation strategy (filled markers) and single continuous time-series (hollow markers) of equivalent total length $\Ne \Naug$. All panels plot results versus $\Naug$ (number of augmented series), where each augmented series contains $\Ne$ events. Top row: $\lambda_0$; middle row: $\alpha$; bottom row: $\delta$. Left column: estimated parameter values, where horizontal dashed lines indicate true values ($\lambda_0=1$, $\alpha=3$, $\delta=10$), markers represent means over 30 trials, and error bars show standard deviations. Center column: relative error between the two approaches, defined as $|\hat{\theta}_{\rm full} - \hat{\theta}_{\rm aug}|/\theta_{\rm true}$, where $\hat{\theta}_{\rm full}$ and $\hat{\theta}_{\rm aug}$ are parameter estimates from the full series and augmented strategy, respectively. Right column: lengths of 95\% confidence intervals, normalized by parameter values to facilitate comparison across different scales.}
    \label{fig:hawkesnaugexp}
\end{figure}
%%%%%%%%%%%%%%%%%%%%%%%%%%%%%%%%%%%%

We wish to test the impact of our proposed data augmentation strategy on parameter estimation.  To do so, we generate a collection of time series each consisting of $\Ne$ events taken from an arbitary time interval in a Hawkes process\footnote{This mimics the situation often found in real-world datasets, where full histories are rarely available and only segments from a limited time interval are observed.}. We compare the best-fit estimators obtained from augmentation of $\Naug$ of these time series against the best-fit parameters derived from a single longer time series with a total of $\Naug \Ne$ events. Fig.~\ref{fig:hawkesnaugexp} illustrates the effect of varying the number of augmentations while holding the number of events per series fixed at $\Ne=16$.

As the figure shows, the data augmentation approach can recover parameters comparable to those obtained from an uninterrupted time series of equivalent total length. That is, our method can yield robust parameter estimates, effectively compensating for sparsity.

%The data augmentation approach can recover parameters comparable to those obtained from an uninterrupted time series of equivalent lengths. This result indicates that, by aggregating multiple smaller time series, our method effectively compensates for the sparsity of each series, yielding robust parameter estimates. 

%%%%%%%%%%%%% SECTION %%%%%%%%%%%%%%%%%%%%
\section{Results for Real-world Data}
\label{sec:Results}

\subsection{Data}

%%%%%%%%%%%%%%%%%%%%%%%%%%%%%%%%%%%%
\begin{figure}[t]
    \centering
    {\includegraphics[width=8.5cm]{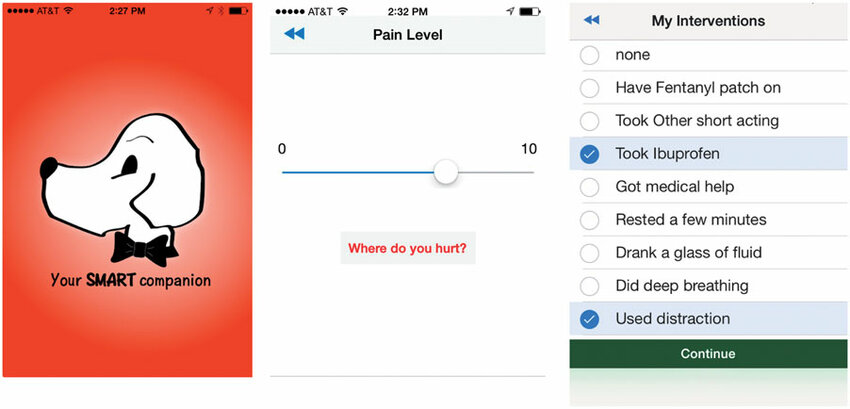}}
    {\includegraphics[width=2.8cm]{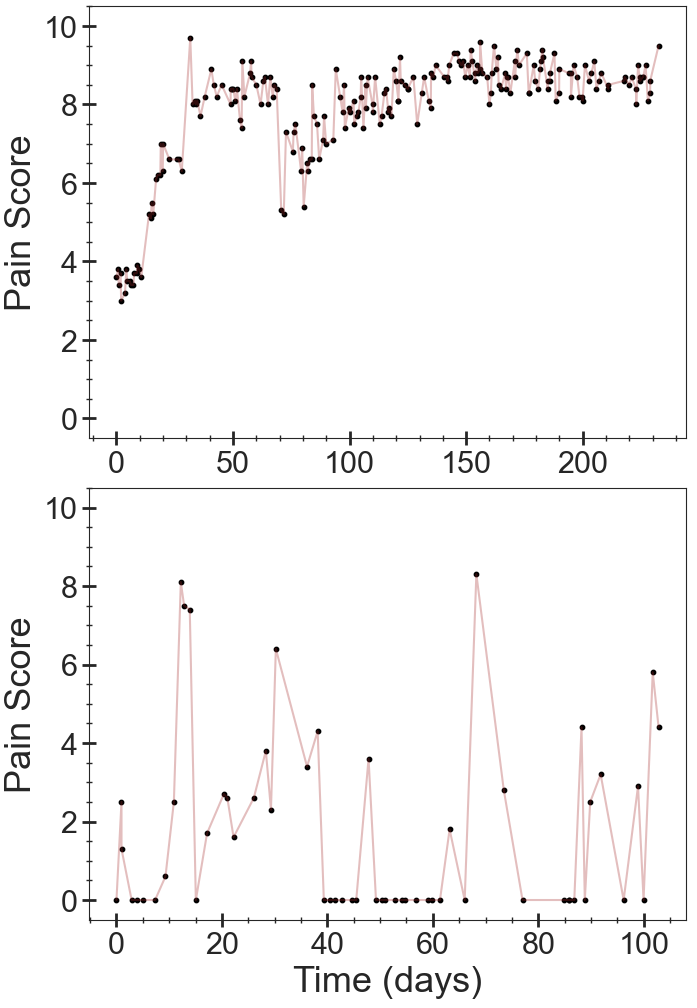}}    
    \caption{\textbf{Real patient data.} Sample screenshots from SMART app and typical patient time-series data collected using the app \cite{shah2014patients}.  Note inter-patient variability, temporal irregularity, reporting fatigue, and other data quality challenges.} 
    \label{fig:garydata}
\end{figure}
%%%%%%%%%%%%%%%%%%%%%%%%%%%%%%%%%%%%

Sickle cell disease (SCD) is a lifelong genetic disorder that affect hemoglobin, which is a carrier of oxygen in red blood cells (RBCs). In SCD, RBCs deform into ``sickle'' shapes, obstructing regular blood flow and causing potentially life-threatening problems. SCD affects more than 100,000 people in the US and 8 million people globally \cite{piel2017global,hassell2010population}. About 90\% of acute care visits for SCD patients are associated with severe and frequent pain episodes. Understanding and modeling these pain episodes is crucial for improving patient care and treatment strategies, as they significantly impact both quality of life and healthcare costs \cite{platt1991pain}. 

Our study employs data from our self-developed Sickle cell Mobile Application to Record symptoms via Technology, or SMART application \cite{shah2014patients, jonassaint2014patients, shah2019digital}. Fig.~\ref{fig:garydata} shows two examples of patient-reported subjective pain data collected via this app. This data comes from a small cohort of 39 patients who were asked to report their pain levels every day. Although subjective pain reports may not fully correspond to physiological indicators, they remain central to SCD pain management because pain is inherently subjective and self-report is currently the only validated method for assessing pain severity (and has been shown to be a reliable indicators of clinical outcomes in SCD \cite{smith2008daily, nasem2020sickle, nasem2025sickle, stewart2021pain}). Among other things, the dataset includes pain levels and corresponding timestamps. 

For our analysis, we treat the event times as the timestamps corresponding to only non-zero pain levels, where each reported pain level above zero constitutes an event. We assume that pain is effectively zero between reports and that the occurrence of pain events exhibits stochastic behavior with temporal dependencies. Even though a lack of report on a particular day is assumed equivalent to a non-event, we acknowledge this assumption may introduce some error given potential reporting fatigue or missed entries. 

%Inconsistent reporting along with the limited duration of the study renders individual datasets sparse and irregular. The sparsity, in particular, can easily lead to inconclusive results when comparing models; this motivated our exploration of new techniques to exploit multiple datasets to gain more statistical confidence.
%. Before we introduce our approach to such data augmentations, we introduce our candidate model to describe the stochastic behavior of pain dynamics. 

\subsection{Model Fit and Distinction}
%%%%%%%%%%%%%%%%%%%%%%%%%%%%%%%%%
\begin{figure}[t]
    \centering
    \includegraphics[width=0.65\columnwidth]{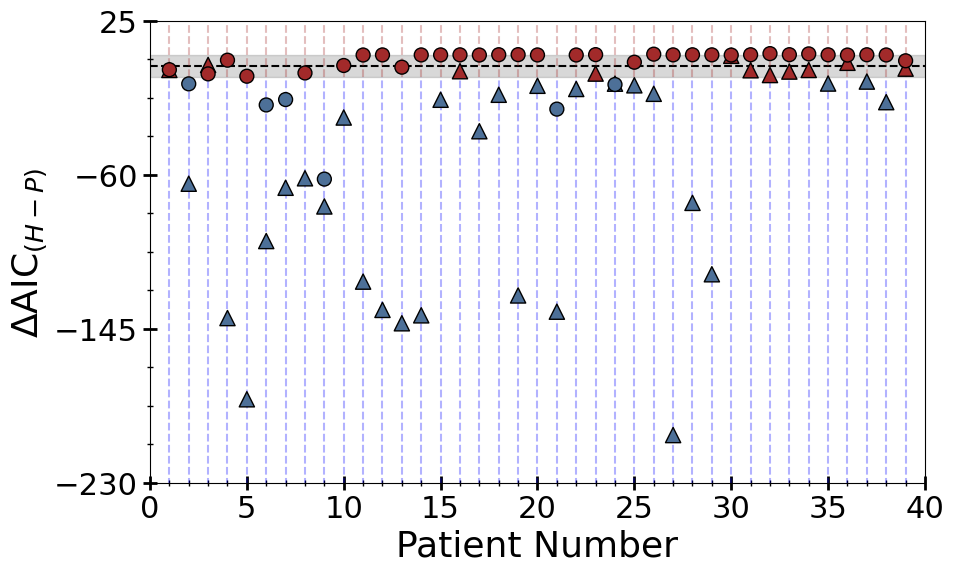}
    \caption{\textbf{Augmentation method applied to patient data.} We plot the difference in AIC (Hawkes minus Poisson) for each time series before (circles) and after (triangles) augmentation. Grey band corresponds to the inconclusive zone (red markers), the region below the band indicates a preference for the Hawkes model with at least 95\% confidence (blue markers). See the right panel of Fig.~\ref{fig:kstest} for the KS-based similarity matrix used in augmentation.}
    \label{fig:realdata}
\end{figure}
%%%%%%%%%%%%%%%%%%%%%%%%%%%%%%%%%

%We now use the data augmentation method to fit candidate Hawkes model and null Poisson model to real-world datasets from the SCD patient cohort.  

In Fig.~\ref{fig:realdata}, we show the results of using our data augmentation method to compare Hawkes and Poisson (null) models for real-world datasets from the SCD patient cohort.  We illustrate the model comparison via $\Delta$AIC both without (circles) and with (triangles) data augmentation.  Color indicates confidence: \textbf{red} for inconclusive regions where $\Delta$AIC lies between $-$6 and 6, and \textbf{blue} for Hawkes fits with more than 95\% confidence ($\Delta$AIC $<-6$).\footnote{Note that the maximum $\Delta$AIC is +6 since the Hawkes model reduces to the Poisson model with appropriate parameter choice, and there is a difference of three in the number of degrees of freedom.}
%\dma{low priority: change color scheme to red and a lighter blue---I don't think these two are easy to distinguish}\kut{Done}

While the single-series fits exhibit a preponderance of cases with preference for the Poisson null model over the Hawkes model (28 of 39), the augmented fits demonstrate a preference for the Hawkes model in an overwhelming majority of cases (36 of 39)---see Table \ref{table:finalresults}.  Notably, Hawkes model selection for augmented datasets occurred even in many cases where the non-augmented dataset had $\Delta$AIC $\approx$ 6, the strongest case for the Poisson null model. 

% The right panel of Fig.~\ref{fig:kstest} shows the KS-based similarity matrix, and Table \ref{table:finalresults} provides a detailed breakdown of patient preferences for the Poisson and Hawkes models across three confidence levels:
% \vspace{-\baselineskip}
% \begin{itemize}
%     \item General preference (|$\Delta$AIC| $>$ 0): Captures all cases where one model is favored over the other.
%     \item Strong preference (|$\Delta$AIC| $>$ 6): Reflects cases with a relative likelihood corresponding to at least 95\% confidence.
%     \item Very strong preference (|$\Delta$AIC| $>$ 9.2): Indicates cases where one model is favored with over 99\% confidence.
% \end{itemize}
% \vspace{-\baselineskip}
% The table shows a marked increase in the number of patients preferring the Hawkes process at both 95\% and 99\% confidence levels after augmentation, underscoring the efficacy of the introduced strategy in amplifying model distinction.  

These results provide compelling evidence for the advantage of leveraging collective likelihoods based on patient similarity in enhancing both model selection and parameter estimation. The observed increase in preference for the Hawkes model aligns well with the hypothesis that temporal dependencies play a crucial role in the data.  %, which may be obscured in single-series analyses.

%%%%%%%%%%%%%%%%%%%%%%%%%%%%%%%%%
\begin{table}[t]
    \centering
        \begin{tabular}{ |c|c|c|c|c| } 
            \hline
            \textbf{Series Type} & \textbf{Confidence Level} & \textbf{Corresponding |$\Delta$AIC|} & \textbf{\# Poisson Patients} & \textbf{\# Hawkes Patients} \\
            \hline
            \hline
            \hline
            \multirow{3}{*}{Single} 
                     & \multicolumn{1}{c}{Basic Preference} 
                     & \multicolumn{1}{c}{0} & \multicolumn{1}{c}{28} & \multicolumn{1}{c|}{11}
                     \\\cline{2-5}
                     & \multicolumn{1}{c}{0.05} & \multicolumn{1}{c}{6}
                     & \multicolumn{1}{c}{2}
                     & \multicolumn{1}{c|}{6}
                     \\\cline{2-5}
                     & \multicolumn{1}{c}{0.01} 
                     & \multicolumn{1}{c}{9.2} 
                     & \multicolumn{1}{c}{0} & \multicolumn{1}{c|}{6}
                      \\
            \hline
            \hline
            \multirow{3}{*}{Augmented} 
                     & \multicolumn{1}{c}{Basic Preference} 
                     & \multicolumn{1}{c}{0} & \multicolumn{1}{c}{3} & \multicolumn{1}{c|}{37}
                     \\\cline{2-5}
                     & \multicolumn{1}{c}{0.05} & \multicolumn{1}{c}{0}
                     & \multicolumn{1}{c}{0}
                     & \multicolumn{1}{c|}{29}
                     \\\cline{2-5}
                     & \multicolumn{1}{c}{0.01} 
                     & \multicolumn{1}{c}{9.2} 
                     & \multicolumn{1}{c}{0} & \multicolumn{1}{c|}{27}
                      \\
            \hline
        \end{tabular}
    \caption{Comparison of AIC-based model preferences: single series fit vs augmented fit. ``Basic Preference'' indicates the model with the lower AIC.
    Confidence level 0.05 reflects a strong preference for one model (|$\Delta\text{AIC}| \gtrsim 6$), and confidence level 0.01 reflects a very strong preference for one model (|$\Delta\text{AIC}| \gtrsim 9.2$).
    %\dma{Since maximum Delta AIC is 6, leading to exp(-6/2) = 0.049 as the smallest p value, I would have expected at least one augmented dataset to prefer Poisson at 0.05 level.  It didn't happen for any of those 3 with basic preference?}\kut{It did not, checked again.}
    }
    \label{table:finalresults}
\end{table}
%%%%%%%%%%%%%%%%%%%%%%%%%%%%%%%%%

%%%%%%%%%%%%%%% SECTION %%%%%%%%%%%%%%%%%%
\section{Discussion}
%\kut{Alternate models: Hawkes with amplitude, continuous models, maybe existing method on clustering, etc}

We have introduced a data augmentation strategy that leverages statistical similarity among sparse time series to improve model selection and parameter estimation---a challenge arising across many scientific domains where individual observations are limited but temporal dependencies are expected.

\textbf{Methodological Contributions.} The augmentation approach addresses two key challenges in analyzing sparse event data. First, it enables model discrimination when individual time series lack sufficient events for conclusive selection. Second, it provides robust parameter estimates by treating statistically similar datasets as multiple realizations of the same underlying process. The collective likelihood framework (Eq.~\ref{collectivelikelihood}) is general and applicable beyond point processes or any specific domain.

%A key insight is the near-flatness of the likelihood surface along curves of constant branching ratio $\alpha/\delta$ in Hawkes processes (Appendix~\ref{app:identifiability}). This identifiability issue is inherent to estimation with limited data \cite{laub2021elements}. 

Our augmentation approach improves precision by increasing the effective sample size while preserving temporal structure, without requiring assumptions about the specific form of dependencies.

% \dma{perhaps cut} We expect this method to be adaptable to any domain with datasets of sparse, clustered events: ecological systems with rare observations, financial markets with infrequent transactions, seismology with incomplete aftershock sequences \cite{ogata1988statistical}, neuroscience with sparse spike trains, or social networks with episodic interactions \cite{rizoiu2021hawkes}. Each case benefits from pooling information across similar observational units. \dma{need citations or need to cut examples}

\textbf{Key limitation.} For the augmentation method to be applicable, \textbf{there must be a reasonable expectation that multiple datasets came from the same (or very similar) models}. This is a key and strong limitation to its applicability.  In an example where time series come from different experiments with vastly different model parameters (or different governing equations altogether), this method would fail and, if employed, could provide erroneous and misleading statistical support for one or more models. 

In the case of SCD, experts in the field typically classify patients into a limited number of categories \cite{Shah2020SCDSeverity, Ballas2010SCDPhenotypes}, lending weight to the convenient modeling assumption that they can be clustered based on dynamics.  

\textbf{Additional limitations.} The KS-based similarity assessment may not capture all temporal structure. Non-transitivity (Fig.~\ref{fig:kstest}) allows augmented datasets to differ across units but introduces potential selection bias. More sophisticated clustering incorporating domain-specific covariates could refine grouping \cite{gelman2013bayesian, raudenbush2002hierarchical}.

Different Hawkes parameters can produce similar interarrival distributions if branching ratios match (Appendix~\ref{app:identifiability}). External covariates could help resolve this degeneracy. The exponential kernel assumes a single memory timescale; many processes involve multiple scales. Constant baseline intensity ignores periodic patterns or trends. Extensions incorporating time-varying parameters, compound kernels, or covariates could address these while preserving the core strategy \cite{reinhart2018review, chiang2022hawkes}.

Computational costs scale as $O(m^2)$ for similarity testing and $O(mk)$ for optimization, where $m$ is the number of units and $k$ is average group size. For large datasets, approximate methods or hierarchical clustering could improve scalability.

\textbf{Comparison with Existing Approaches.} Traditional approaches to sparse data include bootstrapping \cite{efron1994bootstrap}, which generates synthetic observations, and Bayesian methods \cite{gelman2013bayesian}, which incorporate prior information. Our approach differs fundamentally: rather than augmenting individual datasets with synthetic or prior-based data, we pool real observations across statistically similar units. This preserves the empirical nature of inference while increasing effective sample size.

Hierarchical models \cite{raudenbush2002hierarchical, goldstein2011multilevel} also pool information across related groups but require explicit nested structure and shared parameter assumptions. Our similarity-based approach is more flexible, allowing non-hierarchical grouping based on empirical distributional properties. The non-transitivity of similarity (Fig.~\ref{fig:kstest}) means each unit can be augmented with a different subset of the collection, enabling heterogeneous pooling not possible in standard hierarchical frameworks.

For point processes specifically, most augmentation strategies focus on spatial pooling or assume homogeneity across units \cite{reinhart2018review}. Our temporal similarity assessment via interarrival distributions provides a principled criterion for identifying poolable units without requiring spatial structure or homogeneity assumptions.

\textbf{Application to SCD Pain Dynamics.} Our application to sickle cell disease pain events demonstrates practical utility in a real-world clinical context. The shift from 15\% to 74\% of patients showing confident support for a self-exciting process model has implications for management.  It suggests, e.g., that treatment could be improved by enhanced monitoring during high-risk periods following acute episodes, with duration dictated by the memory timescale ($\delta^{-1}$) (which we found to range from 30s to 6 minutes in our data---suggesting risk should return to baseline within about 30 minutes).

%\dma{what is the memory timescale's mean value? distribution? maybe put mean here and histogram for these confident support cases in SI?}.  

Current guidelines emphasize reactive treatment \cite{nasem2020sickle, nasem2025sickle}, but temporal dependencies suggest that interventions preventing initial events or breaking excitation cycles during vulnerable periods may be more effective. The branching ratio $\alpha/\delta$ quantifies self-excitation strength—patients with higher ratios may benefit from aggressive early intervention to prevent cascades, enabling personalized protocols based on individual temporal dynamics \cite{smith2008daily, vuong2023sickle}.

%The application also reveals domain-specific challenges. Pain severity varies but is not captured in our binary framework. The transition to chronic pain states \cite{carroll2018detecting, darbari2016novel} suggests time-varying dynamics not addressed by stationary models. These point to extensions, though, rather than fundamental limitations. \dma{delete par and move refs to future directions section}

\textbf{Future Directions.} Immediate next steps include validation on independent SCD cohorts to assess generalizability, and application to other temporal event datasets where ground truth is known (e.g., simulated epidemic data with known self-exciting parameters). For SCD, integration of clinical covariates (hemoglobin levels, genotype, treatment) into the similarity assessment may improve patient grouping beyond interarrival times alone. 

Extensions to marked point processes (see, e.g., \cite{schlather2001second} or \cite{lotwick1982methods}) could incorporate event severity, addressing a key limitation in the SCD application where pain intensity varies. Time-varying Hawkes models \cite{chiang2022hawkes} combined with our strategy could capture transitions between acute and chronic pain states \cite{carroll2018detecting, darbari2016novel}. Each extension maintains the core principle: leveraging similarity to overcome individual data sparsity.

In the work we present here we have manually selected appropriate dataset augmentation thresholds $p_c$ for each numerical experiment.  Though clearly of interest, we defer for future work the challenge of automatically determining a reasonable threshold, which is connected to the problem of clustering / community detection on a weighted network (the analogue of our p-value matrix $\mathbf{P}$). 

Finally, we have presented our method in the context of selecting among two point process models, but we believe it could be adapted for selection among more than two candidate models and also for continuous time mechanistic models (e.g., dynamical systems), though AIC may need to be employed with cautioin in such cases \cite{utkarsh2025AIC}.

%%%%%%%%%%%%%%% SECTION %%%%%%%%%%%%%%%%%%
\section{Conclusions}

This study introduces a data augmentation strategy for temporal event modeling that addresses challenges posed by sparse individual time series. By pooling statistically similar datasets through collective likelihoods, the approach enables reliable model selection and robust parameter estimation when individual units contain insufficient events for conclusive inference.

We demonstrate the use of this method in the context of pain event data for a collection of 39 patients with sickle cell disease.  The method's applicability ultimately rests on a key assumption that multiple sparse datasets originated from the same (or nearly the same) model.  In situations where this is plausible, we expect our framework to enable reliable inference from fragmented data, advancing our ability to understand and predict the dynamics of complex systems. 

%The method's applicability stems from three features \dma{these are particular to point processes, but models could go beyond that (e.g. including earthquake magnitudes)}: it requires only interarrival time distributions for similarity assessment; it preserves temporal structure while increasing effective sample size; and it integrates with standard likelihood-based inference. Validation on sickle cell disease pain events demonstrates practical utility. We expect the framework to be relevant to problems with sparse event data arising in many domains of science and engineering. By enabling reliable inference from fragmented data, this work advances our ability to understand and predict the dynamics of complex systems.

\enlargethispage{20pt}

%\dma{please copy the below statements from examples in recently published RSIF papers}

% \ethics{Insert ethics text here.}

\dataccess{The data used in the Results section is shared along with the manuscript.}

\aucontribute{K.U. and D.M.A.: conceptualization, formal analysis, methodology, visualization, writing---original draft, writing---review and editing. N.R.S.: clinical guidance, data. T.B: mathematical and computational guidance. All authors gave final approval for publication and agreed to be held accountable for the work performed therein.}

\competing{K.U. and D.M.A. declare no conflict of interest in this study.}

\funding{This study was supported by NIH grant no. 5R01AT010413.}

\ack{The authors are grateful to Dr. Gary K. Nave Jr. (Colorado School of Mines) and Richard Suhendra (Northwestern University) for their helpful discussions.} 

% \disclaimer{Insert disclaimer text here.}

\printbibliography

\newpage \clearpage
\appendix

\section{Stationary Mean Intensity of a Hawkes Process} 
\label{app:hawkes_equil}

Consider a Hawkes process with baseline intensity $\lambda_0$ and an exponential kernel
\begin{equation*}
    \Phi(t) = \alpha e^{-\delta t}, \quad t \ge 0.
\end{equation*}
The conditional intensity at time $t$ is given by
\begin{equation*}
    \lambda(t) = \lambda_0 + \sum_{i:t_i < t} \Phi(t - t_i) 
                = \lambda_0 + \alpha \sum_{i:t_i < t} e^{-\delta (t - t_i)},
\end{equation*}
where the sum is over all events that occurred before time $t$. Each past event contributes to the current intensity, with a contribution that decays exponentially over time.

\textbf{Branching process interpretation.} 
One way to look at the Hawkes process is through its branching structure. Each event can be thought of \cite{laub2021elements} as generating future ``offspring’’ events independently according to a Poisson process with mean
\begin{equation*}
    \eta = \int_0^\infty \Phi(t) \, dt = \frac{\alpha}{\delta}.
\end{equation*}
This quantity $\eta$ is called the \emph{branching ratio}. If $\eta < 1$, the process is subcritical, meaning that on average each event produces less than one descendant. This ensures that the process does not explode and admits a stationary distribution.

\textbf{Stationary mean intensity.} 
Let $\lambda_c = \mathbb{E}[\lambda(t)]$ denote the stationary mean intensity. By stationarity, the expected total intensity can be decomposed into the baseline intensity plus the expected contribution from all offspring events. In other words, each event contributes to future intensity on average according to the branching ratio:
\begin{equation*}
    \lambda_c = \lambda_0 + \eta \, \lambda_c.
\end{equation*}
Here the first term $\lambda_0$ accounts for the spontaneous events, while the second term $\eta \, \lambda_c$ accounts for the self-excitation of the process in expectation. Solving for $\lambda_c$ gives the exact stationary mean intensity:
\begin{equation}
    \lambda_c = \frac{\lambda_0}{1 - \eta} = \frac{\lambda_0}{1 - \alpha / \delta}.
\end{equation}

The branching ratio $\eta$ provides an intuitive measure of how strongly the process self-excites: as $\eta \to 1$, the stationary mean diverges, reflecting the transition to a critical process.

\section{Identifiability of \texorpdfstring{$(\alpha,\delta)$}{alpha/delta} in the Exponential Hawkes Process}
\label{app:identifiability}

\begin{figure}
    \centering
    \includegraphics[width=0.5\linewidth]{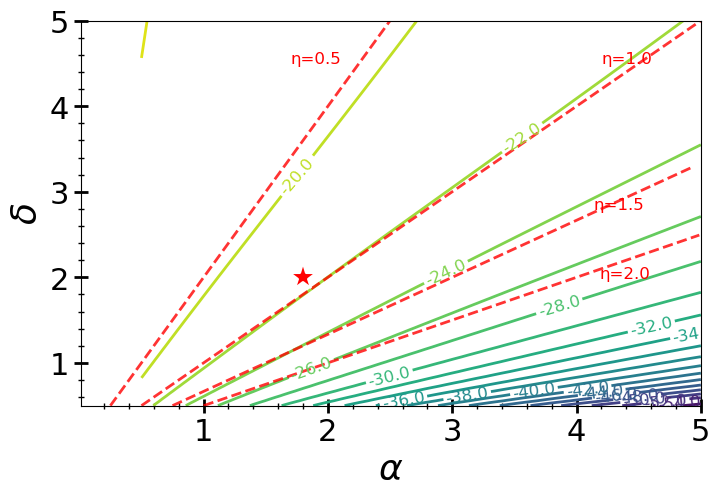}
    \caption{\textbf{Likelihood ridge along constant branching ratio reveals practical non-identifiability.} Practical non-identifiability of $(\alpha,\delta)$ in the exponential Hawkes process. Log-likelihood surface for synthetic data with $N=6$ events over $[0,10]$, showing contours and lines of constant branching ratio $\eta = \alpha/\delta$ (red dashed). The true parameters ($\alpha=1.8$, $\delta=2.0$, red star) lie on the $\eta=0.9$ ridge. Likelihood contours run nearly parallel to constant-$\eta$ lines, creating a flat direction. The numerical evaluation shows log-likelihood varies by only $\Delta \log L\approx 0.8$ across a wide range of $\alpha$ values for $\eta=0.9$, demonstrating a relatively weaker constraint on individual parameters despite a strong constraint on their ratio. This flatness explains why MLE optimization often yields parameter estimates that converge close to, but not exactly at, the true values.}
    \label{fig:optlandscape}
\end{figure}

In Fig.~\ref{fig:hawkesnaugexp}, we note that the estimated parameters converge nearly---but not exactly---to the true values; this is a result of a near-failure of parameter identifiability for the Hawkes model.  True failure would occur if one or more likelihood contours in the $(\alpha,\delta)$-plane became straight lines (see Fig.~\ref{fig:optlandscape}).

%Notably, we observe that the estimates for $\lambda_0$, $\alpha$, and $\delta$ remain stable and within acceptable error margins across a range of augmentation scenarios. We also observe the errors consistently decreasing with increasing $\Naug$. This consistency highlights the potential of the augmentation strategy to provide reliable parameter estimates even when dealing with fragmented or limited observational data.

%We give a brief explanation for the practical non-identifiability of the parameters $(\alpha,\delta)$ in the exponential Hawkes process.  

The issue is not that the model is theoretically unidentifiable, but that in finite samples (especially with short observation windows or few events) the likelihood surface becomes nearly constant along certain curves in the $(\alpha,\delta)$-plane, making different parameter pairs yield almost indistinguishable
fits.

Consider the Hawkes process with exponential memory kernel
\begin{equation}
    \Phi(t) = \alpha e^{-\delta t}, \qquad \alpha>0,\ \delta>0,
\end{equation}
and intensity
\begin{equation}
    \lambda(t) = \lambda_0 + \sum_{i:t_i<t} \alpha e^{-\delta (t - t_i)}.
\end{equation}
For event times $\{t_i\}_{i=1}^N$ on $[0,T]$, the log-likelihood is
\begin{equation}
\mathcal{L}(\alpha,\delta)
= 
\underbrace{\sum_{i=1}^N \log \lambda(t_i)}_{\text{log–intensity at events}}
\;-\;
\underbrace{\int_0^T \lambda(t)\, dt}_{\text{compensator}}.
\end{equation}

For the exponential kernel, the compensator (the integrated intensity function) has the closed form
\begin{equation}
    \int_0^T \lambda(t)\,dt
    = \lambda_0 T
      + \frac{\alpha}{\delta}
        \sum_{i=1}^N\!\left(1 - e^{-\delta (T - t_i)}\right).
    \label{eq:hawkes-integral}
\end{equation}
When both the decay rate $\delta$ is sufficiently large and events occur sufficiently early in the observation window such that $\delta(T-t_i) \gg 1$ for most events, the terms satisfy $e^{-\delta(T-t_i)}\approx 0$ \dma{shouldn't this apply for large $\delta$? but it seems like you're suggesting it applies for small $\delta$} \kut{Corrected again}.  In this regime,
\begin{equation}
    \int_0^T \lambda(t)\,dt
    \approx \lambda_0 T + \frac{\alpha}{\delta} N,
    \label{eq:hawkes-integral-approx}
\end{equation}
so the integral depends almost entirely on the branching ratio
$\eta = \alpha/\delta$.  This produces an exactly linear contour in the likelihood plane coming from the compensator alone.

The intensity at an event time is
\begin{equation}
    \lambda(t_i)
    = \lambda_0 + \sum_{j<i} \alpha\, e^{-\delta (t_i - t_j)}.
\end{equation}

The contribution of each past event involves the product $\alpha e^{-\delta\Delta t}$, where $\Delta t = t_i - t_j$.  In principle this term distinguishes $\alpha$ and $\delta$, since it does not collapse to a function of $\alpha/\delta$.  However, the excitation contributions $\sum_{j<i} \alpha e^{-\delta(t_i - t_j)}$ become negligible compared to $\lambda_0$ when either (i) the excitation is weak ($\alpha/\lambda_0 \ll 1$), making the process only weakly self-exciting, or (ii) the decay rate $\delta$ is large relative to typical inter-event times, causing excitations from past events to die out before the next event occurs.  In these regimes, $\lambda(t_i) \approx \lambda_0$ for most events, so the log-intensity contributions $\log\lambda(t_i)$ vary only weakly with $(\alpha,\delta)$.  Consequently the event-time term cannot significantly bend the nearly linear contours produced by the compensator.

\dma{I think the argument is just that the sum term is much smaller than the integral term in some limit but I'm not sure what the limit is.  I don't see how the sum term on its own would ever produce a linear contour.}\kut{Rewrote} \dma{I still don't follow the argument in these three cases but if you're confident I'm okay with submitting and we can discuss later (given that this is appendix material)}

\kut{Clarified that compensator approximation requires both large $\delta$ and early events (joint condition), while sum term weakness requires either small $\alpha/\lambda_0$ or large $\delta$ (alternative conditions).}

\noindent Combining the two effects:
\vspace{-\baselineskip}
\begin{itemize}
    \item the compensator term is nearly invariant along curves of constant $\alpha/\delta$,
      and
    \item the event-time intensities only weakly constrain $(\alpha,\delta)$ unless the data
      contain inter-event gaps that probe the exponential decay scale,
\end{itemize}
\vspace{-\baselineskip}
the log-likelihood develops a nearly constant ridge along curves close to
\begin{equation}
    \{(\alpha,\delta): \alpha/\delta \approx \text{constant}\}.
\end{equation}
This explains why parameter estimates often converge close to --- but not exactly equal to—the true values unless large
datasets or very tight optimization tolerances are available. We can visualize the ridges in Fig.~\ref{fig:optlandscape}.

\section{Small-Sample Correction for Model Selection}
\label{correctedAICmodelcomp}

\begin{figure}[t]
    \centering
    \includegraphics[width=0.6\columnwidth]{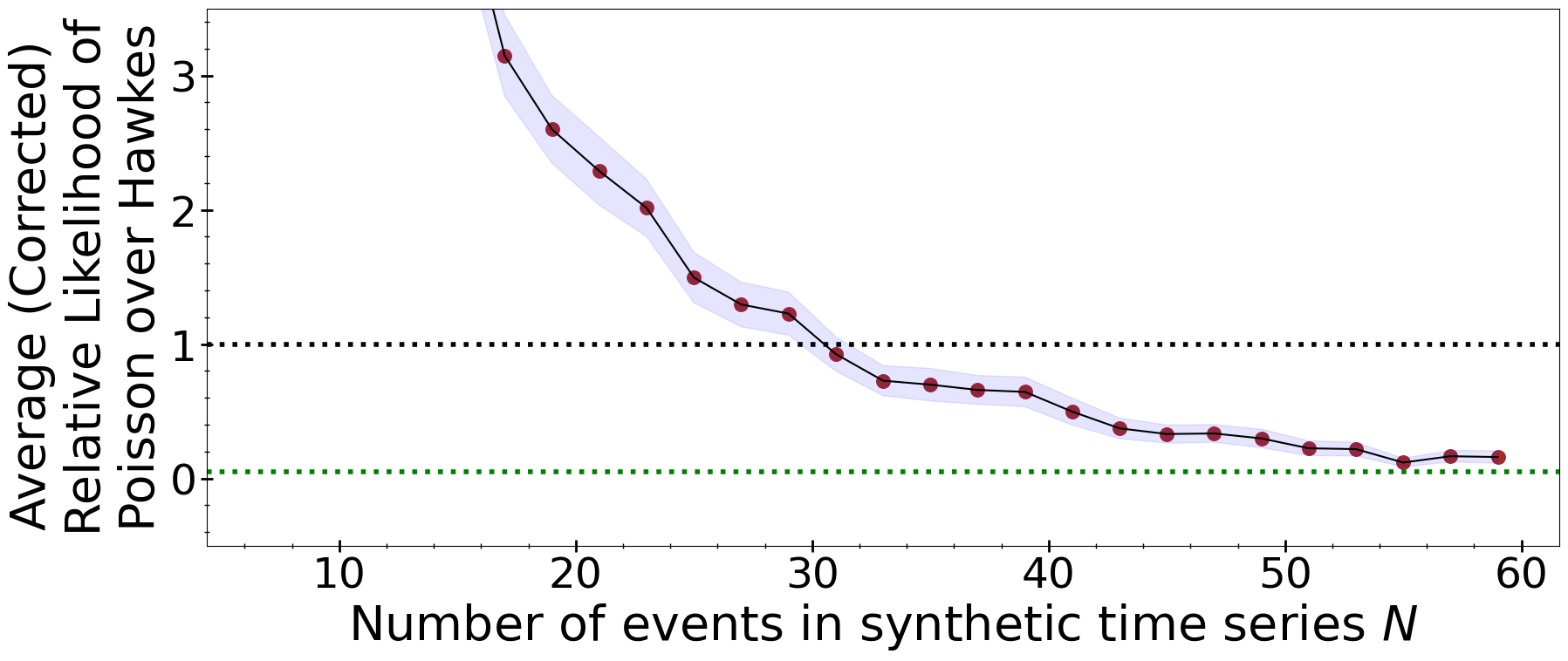}
    \caption{\textbf{AICc correction does not resolve sparse-data model selection challenges.} Number of data points needed to distinguish the Hawkes model from the Poisson model using AICc. As in Fig.~\ref{fig:modelcomp}, black dashed line shows basic preference ($\mathcal{L}=1$), green dashed line shows 95\% confidence ($\mathcal{L}=0.05$), red markers indicate averages over 50 repetitions, and purple shading denotes the 95\% confidence interval. Parameter choices: ($\lambda_0, \alpha, \delta$)=(1, 3, 6).}
    \label{fig:modelcomp_aicc}
\end{figure}

To assess the impact of small datasets on model selection, we recomputed the model comparison using the corrected Akaike Information Criterion (AICc) \cite{Hurvich1989}. Figure~\ref{fig:modelcomp_aicc} shows results analogous to Fig.~\ref{fig:modelcomp}, illustrating how AICc performs almost equivalently to AIC qualitatively.

%%%%%%%%%%%%%%%%%%%%%%%%%%%%%%%%%
%%%%%%%%%%%%%%%%%%%%%%%%%%%%%%%%%

\end{document}